\documentclass[12pt,a4paper]{article}
\usepackage{graphics}
\usepackage{array}
\usepackage{hhline}
\usepackage{subfigure}
\newcommand{\be}{\begin{equation}}
\newcommand{\ee}{\end{equation}}
\newcommand{\bea}{\begin{eqnarray}}
\newcommand{\eea}{\end{eqnarray}}
\newcommand{\nn}{\nonumber\\}
\begin{document}
\title{Quark Initial State Interaction in Deep Inelastic Scattering and the
  Drell-Yan process
\thanks{work supported by DFG and BMBF}}
\author{O.~Linnyk\thanks{olena.linnyk@theo.physik.uni-giessen.de} ,
  S.~Leupold, U.~Mosel}
%
%\affiliation
%{Institut fuer Theoretische Physik, Universitaet Giessen,
%  Heinrich-Buff-Ring, 16, D-35392 Giessen, Germany}
%
\date{\small Institut f\"ur Theoretische Physik, Universit\"at Giessen, Germany \\
 \today}
\maketitle
\begin{abstract}
We pursue a phenomenological study of higher twist effects in high energy
processes by taking into account the off-shellness (virtuality) of partons 
bound in the nucleon.
The effect of parton off-shellness in deep inelastic $ep\to eX$ scattering (DIS) and the Drell-Yan 
process ($pp\to l\bar l X$) is examined. Assuming factorization and a
single-parameter Breit-Wigner form for the parton spectral function, we develop a model to calculate the
corresponding off-shell cross sections. Allowing for a finite parton 
width $\approx 100$~MeV, we reproduce the data of both DIS and the triple differential Drell-Yan cross section 
without an additional K-factor. The results are compared to those from perturbative QCD and the intrinsic-$k_T$ approach. 
\end{abstract}
%
%\PACS{{13.60.Hb}{Total and inclusive cross sections (including deep-inelastic processes)} \and {13.85.Qk}{Inclusive production with identified leptons, photons, or other nonhadronic particles}} 
%
%%%%%%%%%%%%%%%%%%%%%%%%%%%%%%%%%%%%%%%%%%%%%%%%%%%%%%%%%%%%%%%%%%%%%
\section{Introduction}

\label{intro}
 
One of the major goals of present day research is to study the
structure of the nucleon and other hadrons in terms of the
fundamental quark-gluon dynamics.
In high energy hadronic processes like DIS, 
the Drell-Yan process, jet production, {\it etc.,} the soft and 
hard subprocesses can be disentangled. 
The hard cross section can be calculated using the well established
methods of perturbative QCD.
This procedure 
allows one to extract 
the information about the nonperturbative quark and gluon 
properties in a bound state from the experimental data. 

The described method, based on factorization, is analogous to the 
Plane Wave Impulse Approximation (PWIA) for the description of quasi-elastic ($e,e'p$) 
scattering in nuclear physics. The approximation of quasi-free constituents 
is valid when the binding energy is small compared to the energy transfered.
In the theory of nuclei, the effects beyond the PWIA (such as photon 
radiation, initial state interaction (ISI) and final state interaction) 
are known to be essential for understanding se\-mi-\-exclu\-sive observables.
Measurements, in which energy and momentum of the nucleon can be determined 
from the final state kinematics, offer an opportunity to study these effects 
and thus probe the nucleon interaction in nuclei~\cite{benhar.A,paris}.
One would like to gain an understanding of the hadron structure which
is as good as the present understanding of the compositeness of the
nucleus in terms of nucleons and their interaction. 

\begin{figure}
\begin{center}
\resizebox{0.7\textwidth}{!}{%
  \includegraphics{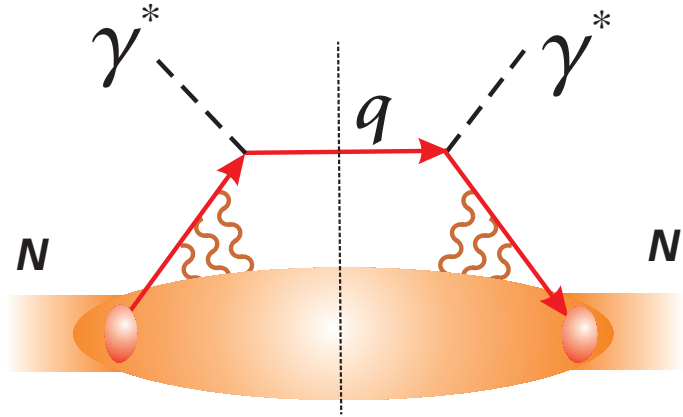}
}
\end{center}
\caption{The handbag graph for DIS and the relevant initial 
state interactions that could build up a finite parton width.}
\label{diagram}
\end{figure}

In the present paper, deep inelastic scattering and the Drell-Yan pair production 
are considered. Our aim is to investigate a kinematical region where 
standard perturbative QCD no longer works and where
we thus need to model nonperturbative effects. 
Higher twists, suppressed 
by inverse powers of the hard scale $Q^2$, are important
in description of low $x_{Bj}$ DIS \cite{brodsky.hoyer}, hadron-hadron
collisions \cite{Qiu}, and semi-inclusive
DIS at moderate energies \cite{Mulders}. 
In the case of fully inclusive DIS, the
factorization of higher twist contribution in terms of a hard coefficient and
the matrix element of quark and gluon fields in the nucleon was proven
\cite{Qiu}. Coefficients of the twist expansion were calculated in
\cite{Vainshtein}. 
But the matrix element is a non-perturbative object and has
to be modelled. 
In the present work we model the power corrections by dressing 
the active parton lines with spectral functions.
Figure \ref{diagram} shows a handbag graph with the relevant 
initial state interactions which could build up a finite parton width.

The initial and final state parton interaction effects 
on the observable hard scattering cross sections have recently attracted a lot 
of attention. 
The essential role of final state interactions in the interpretation of 
the measured DIS structure functions at finite $Q^2$ has been stressed in
\cite{brodsky.hoyer,hoyer1,hoyer2}. 
In several other calculations~\cite{kt,kt0,kt1,kt1a,kt2,kt2a}, non-collinear 
kinematics, {\it i.e.,}~non-vanishing primordial transverse momenta of the 
partons in the nucleon, was considered. 
The authors of \cite{Brodsky1,BrodskySSA} pointed out that one gluon exchange in the
initial state can produce a large effect in $\pi p$ scattering in the framework 
of a quark-diquark model. 
On the other hand, parton off-shellness effects in DIS and the Drell-Yan 
process can have the same order of magnitude as those of the intrinsic transverse 
momentum~\cite{us1}. Thus, a consistent treatment of both is necessary 
to go beyond the PWIA in these reactions. 
In this paper, we develop the formalism to study these effects and apply it to calculate 
the cross sections of DIS and the Drell-Yan process. 

Since the triple differential Drell-Yan cross section is a more
exclusive observable than the DIS cross section, it is expected to be more 
sensitive to the ISI. The results of our calculations confirm this intuitive expectation.
By taking into account both the finite width generated by ISI and the
non-collinearity of partons, we reproduce the experimentally measured fully 
inclusive DIS cross section and the triple-differential cross section of 
the Drell-Yan process very well.
Our success in reproducing the transverse momentum 
distribution of the Drell-Yan lepton pairs is particularly interesting, since 
other models disagree with experiment. In leading order of perturbative 
QCD, a delta function at zero transverse momentum is predicted. 
Only after the resummation of all orders in $\alpha_S$, the pQCD
predictions for the Drell-Yan pair $p_T$-distributions are reliable 
\cite{logM/P,Qiu.Zhang,Landry}.
Among the next-to-leading order contributions to the dilepton cross
section, only the gluon Compton scattering can give non-vanishing
transverse momentum ($p_T$).
This process, however, contributes only
in the region of very high $p_T$: $p_T \ge \sqrt{M}$, where $M$ is
the mass of Drell-Yan pair.
In contrast, the major part of the measured pairs lies in 
the interval $0 \! < \! p_T \! < \! \sqrt{M}$ and is not described by leading
twist perturbative QCD \cite{exp}. 
At the same time, none of the phenomenological models, including intrinsic
$k_T$ approach, is 
able to reproduce simultaneously magnitude and shape of the
experimentally observed distribution. In contrast, as we will
demonstrate below, the data can be successfully described by 
a model which allows for off-shell partons.

The applied technique is presented in section \ref{method} and the obtained 
results are discussed in section \ref{results}, followed by a short summary.
We discuss the interesting question of the application of
factorization to hadron scattering in the Target 
Rest Frame in the Appendix.

%%%%%%%%%%%%%%%%%%%%%%%%%%%%%%%%%%%%%%%%%%%%%%%%%%%%%%%%%%%%%%%%%
\section{Method}
\label{method}

The basic tool in the calculation of hard processes is the factorization into 
hard and soft physics: 
\be 
\label{fact}
d \sigma = \sum _i e_i^2 f_i ( \xi , \vec p_\perp) \otimes 
d \hat \sigma ( \xi, \vec p_\perp),
\ee 
where the sum runs over all relevant parton flavors, $e_i$ is the charge of the 
$i$th type of parton in units of the proton charge $e$.
$d \hat \sigma$ is the elementary cross section for a given process,
$f (\xi , \vec p_\perp)$ are unintegrated parton distributions
defined as~\cite{collins}:
\be
\label{unintegrated.PDF}
f ( \xi , \vec p_\perp) = 
\frac{1}{4\pi } 
\int  d^4 y  \
\langle N | \bar \psi (y) \gamma^+ 
\psi (0) | N \rangle \
e ^{i p \cdot y} \delta (y^+) , 
\ee
%
%\be
%[ y,0 ] \equiv \mbox{P} \exp \left[ \frac{ig}{2} \int _0 ^y d\tau A^+ (\tau ) \right] ,
%\ee
 where $\xi \equiv p^+/P^+$ is the Nachtmann variable and $p$ and $P$ are
 momenta of  the active parton and hadron, respectively.
 In \cite{Ji.fact}, 
the factorization in the form (\ref{fact}) was proven in the leading power 
of the hard scale (photons virtuality in DIS and the Drell-Yan process).

Note that the unintegrated distributions do not depend on $p^-$ due to $\delta (y^+)$. In 
other words, the parton distributions measure the correlation of partons with equal light
cone times ($y^+=0$). This reflects the fact that the structure functions, measured in 
the fully inclusive DIS, depend only on $P^+$ in the Bjorken limit. 
To see this, let us consider the  hadron tensor measured in DIS~\cite{jaffe}:
\be 
\label{exponent}
W_{\mu \nu} (q) = 
\frac{1}{4\pi } 
 \int  d^4 y e ^{iq\cdot y} 
\langle N |J_\mu (y) J_\nu (0) | N \rangle _c . 
\ee 

In the hadron rest frame ($M_N$ is the nucleon mass) :
\be
(q_+,q_-,\vec q_T) = (-M_N x_{Bj},\frac{Q^2}{M_N x_{Bj}},\vec 0),
\ee
As $Q^2 \rightarrow \infty $ with $x_{Bj}\equiv Q^2/P\cdot q$ finite and fixed, 
$q_-\rightarrow \infty$. As a consequence, the integral in (\ref{exponent}) 
should vanish due to the fast oscillating exponent, unless
\be
y _+ \rightarrow 0.
\ee
At the same time, $y_-$ is finite and even large $y_-$ can contribute to 
$W_{\mu \nu}$ in some cases. To be precise, the restriction on $y_-$ is \cite{jaffe} :
\be
|y_-| < 1/(M_N x_{Bj} ).
\ee
In case of a fully inclusive process, one has the following condition due to causality:
\be
y^2=y_- y_+ -\vec{y}^2_T \ge 0 \mbox{ } \Rightarrow \mbox{ } \vec{y}_T \rightarrow \vec 0.
\ee 
Thus, DIS in the Bjorken limit is a light-cone ($y^2 \rightarrow 0$) dominated process and the hadronic 
part of the DIS cross section  is a function of a single 
variable $p^+\equiv x_{Bj}P^+$.

The factorization formula (\ref{fact}) is valid only in the scaling limit,
{\it i.e.}, at the leading power as $Q^2\to \infty$. 
On the other hand, at moderate $Q^2$ considerable $p^-$-dependent corrections 
might be necessary to make 
predictions for semi-exclusive observables, {\it e.g.} Drell-Yan lepton 
pair production cross section and asymmetries.  
In this case, we propose the following factorization ansatz: 
\be 
\label{fact2}
d \sigma = 
\sum _i e_i ^2
g_i ( \xi , \vec p_\perp , p^-) \otimes 
d \hat \sigma (\xi, \vec p_\perp ,p^-).
\ee 

The difference between (\ref{fact}) and (\ref{fact2}) is precisely due to
off-shellness 
effects that we aim to study. Indeed, the minus component of the free parton momentum is
fixed by the on-shell condition:
\be
\label{on-shell_condition}
p^2= p^+ p ~^- -\vec p_{\perp} ^2 = \xi P^+ p^- - \vec p_{\perp} ^2 = 0
\ee
(we put the current quark mass to zero).
However, since the partons are bound in the nucleon, (\ref{on-shell_condition}) 
no longer holds. Thus all the 4 components of the parton momentum are independent
and the full propagator should be used.
In this case, the cross section is calculated using a virtuality distribution
defined by a parton spectral function~\cite{benhar.N,benhar.flux}. Spectral functions of quarks in quark 
matter are, for example, calculated in \cite{froemel}.

In nuclear physics, the terms off-shellness and virtuality are often
interchanged. 
The on-shell condition for the nucleon reads $P^2 = M_N^2$, where
$M_N$ is the nucleon mass in vacuum. Thus, only 3 components of the
on-shell nucleon's 4-momentum are independent. 
In case of an interacting nucleon, $P^2$ is no longer fixed and its distribution 
(spectral function) is given by the details of the interaction. 
All four components of the off-shell nucleon's momentum are independent.
Thus, a hadron is said to be 
off-shell, if its momentum squared is different from the free hadron
mass, {\it i.e.} when it is virtual.

Partons in the nucleon are always virtual. For example, in the naive parton model, the
parton momentum squared is $p^2=(x_{Bj} P)^2=x_{Bj}^2 M_N^2$, which is usually far from 
the current parton mass ($=0$ in our calculations). 
We call a parton off-shell,
if the parton's momentum has four independent components. In this case the
parton off-shellness $p^2$ is not fixed and should be integrated over. 
This differs from the "trivial off-shellness" of parton model, 
in which  the quark is virtual, but it's off-shellness is fixed 
(to $x_{Bj}^2 M_N^2$ ).
More realistically, one should include 
the transverse motion of partons. Then, for a free parton, $p^- = p_\perp
^2/p^+$. In our calculations, $p^-$ is not fixed by $p^+$ and $p_\perp ^2$. 
Instead, we integrate over all kinematically allowed $p^-$.

In the following, we additionally assume that the dependence of $g$ on $p^-$ factorizes from 
the $p_{\perp}$-dependence:

\be 
\label{g_fact}
d \sigma = 
\sum _i e_i^2
\tilde f_i ( \xi ,\vec  p_\perp ) \otimes d \hat
\sigma (\xi , \vec p_\perp, m) \otimes \mbox{A} (m , \Gamma) . 
\ee
In (\ref{g_fact}), $d \hat \sigma(\xi, \vec p_\perp , m)$ is the off-shell partonic cross 
section and $m\equiv \sqrt{p^2}$ the parton's off-shellness.
We choose 
\be
\label{choice_distribution}
\tilde f_i (\xi ,\vec  p_\perp ) = f_i (\xi ,\vec  p_\perp ).
\ee
Identifying $\tilde f_i (\xi, p_\perp)$ with the usual parton distribution
functions means that we apply a quasiparticle picture, in which all
effects involving more than one parton are encoded in the spectral function.
The latter includes a width caused by parton-parton interactions (see {\it
  e.g.} \cite{froemel} and references therein).

In our calculations, a Breit-Wigner parametrization for the parton 
spectral function $\mbox{A}(m,\Gamma)$ was applied: 
\be 
\label{BW}
\mbox{A}(m,\Gamma) = \frac{1}{\pi}\frac{\Gamma}{m^2+\frac{1}{4}\Gamma^2} .
\ee
The width $\Gamma $ of partons was considered constant. We find its value by comparing 
the calculated cross section to the data.
Note that for simplicity we use the same $A(m)$ for all parton types.

The hard part, {\it i.e.,}~the partonic cross section, is calculated using 
the rules of perturbative Quantum Chromodynamics (pQCD). We calculated 
the pQCD differential cross section for an electron scattering off a virtual 
quark and that for the annihilation of an off-shell quark-antiquark pair 
into a pair of dileptons (see (\ref{DY2}) below). 
Both off-shell partonic cross sections turned out to be gauge invariant due to 
the on-shell leptons making the modification of the 
vertex by Ward's identity unnecessary.

The analysis of the off-shell kinematics and the obtained 
cross sections are separately given below for DIS and the Drell-Yan process. 
The case of electron scattering (section \ref{m_DIS}) is 
simpler and serves as an introduction to the calculation of the Drell-Yan 
pair transverse momentum distribution in section \ref{m_DY}.

%%%%%%%%%%%%%%%%%%%%%%%%%%%%%%%%%%%%%%%%%%%%%%%%%%%%%%%%%%%%%
\subsection{DIS}

\label{m_DIS}

Ignoring the off-shellness of partons, the factorization formula (\ref{fact}) 
for DIS can be written as:
\be
\label{fact1}
\frac{d \sigma}{ d \hat t d \hat u }= \sum _i e_i ^2 
\int \limits _0 ^1 d \xi \int d \vec p_{\perp}
 f_i (\xi , p_{\perp}) 
\frac{d \hat \sigma }{ d \hat t d \hat u }  ,
\ee
\be
\left( \frac{d \hat \sigma }{ d \hat t d \hat u } \right)_{\mbox{\small on-shell}} =
\frac{2\pi \alpha ^2}{ \hat t^2  \hat s^2} \left( \hat s^2+ \hat u^2 \right) 
\delta (\hat s +\hat u + \hat t),
\ee
where $s, t, u$ are the Mandelstam variables, $\alpha = e^2 / 4 \pi$, the parton quantities are 
labeled with hats, and the $\delta $-function reflects the on-shell condition
on the parton level: 
\be
\label{conserv}
\hat s+\hat u+\hat t=0.
\ee

Let us consider the Bjorken limit ($Q^2 \rightarrow\infty$ with $x_{Bj}$
-fixed, where $q$ is the momentum transfer, $Q^2\equiv -q^2$) in the
rest frame of the nucleon. In this limit, the partonic and hadronic invariants are
simply related:
\be
\label{hats}
\hat t = t,\mbox{ }\mbox{ } \hat s = \xi s, \mbox{ } \mbox{ } \hat u = \xi u.
\ee
From (\ref{conserv}) and (\ref{hats}) one gets the constraint 
\be
\label{bj}
\xi \rightarrow -\frac{t}{s+u}= - \frac{q^2}{2 P\cdot q} \equiv x_{Bj}.
\ee
The parton model cross section of DIS is obtained:   %restored
\be 
\label{pDIS}
%\frac{d \sigma}{ d Q^2 d x_{Bj} }  
%=
\left( \frac{d \sigma}{ d t d u } \right) _{\mbox{\small LO}}
= 
\sum _i e_i ^2  x_{Bj} f_i ( x_{Bj} ) 
\left( \frac{2\pi \alpha ^2}{ t ^2 s^2} \frac{(s^2+u^2)}{s+u}
\right) ,
\ee
where 
$$f_i (x_{Bj}) \equiv \int d p_{\perp} f_i (x_{Bj},p_{\perp}),$$
''LO'' stands for leading order of perturbative QCD, {\it i.e.} parton model.

For finite $Q^2$, the fact that the partons are off-shell can 
generate large corrections to the
formulas (\ref{conserv})-(\ref{pDIS}). We would like to point out the
important analogies and differences to the on-shell case:  

\begin{itemize}
\item The energy-momentum conservation reads ({\it c.f.}~(\ref{conserv}))
\be
\hat s+\hat u+\hat t= m ^2,
\ee
where $m^2 \equiv p^2$ denotes the virtuality of the struck parton.  
\item In case of an off-shell initial quark, we find the following 
relation between the partonic and hadronic variables 
\bea
\label{off-shell-Mandelstam}
\hat t = t = -Q ^2, \mbox{   } \mbox{   }
\hat s = \xi (s - M_{N}^2) +  m ^2 ,  \mbox{   }\mbox{   } \nn
\hat u = Q^2 -\xi (s-M_N^2)=
\xi  (u + M_{N}^2) +Q ^2(1-\frac{\xi}{x_{Bj}}) ,
\eea
which coincides with (\ref{hats}) in the Bjorken limit. 
We choose the $z$-axis along the incoming electron.
\item The hadron light cone momentum fraction carried by the struck
   parton ($ \xi \equiv  p^+ / P^+ $) is not equal to the Bjorken
   $x_{Bj}$, unless $Q^2 \to \infty$. The relation between $x_{Bj}$ and 
   $\xi$ is
\be
\label{kinDIS}
x_{Bj}=
\xi 
\frac{Q^2
        }{
          Q^2 - m^2 - 
          \xi (M_N^2 -\frac{m^2+ \vec p_\perp ^2}{\xi ^2})\frac{Q^2}{s-M_N^2}+
          2 \vec p_{\perp} \vec q_\perp }.
\ee
Relation (\ref{kinDIS}) yields a nonlinear equation for $x_{Bj}$, because 
$\vec q _{\perp}$ depends on $x_{Bj}$ as follows: 
\be
\vec q _{\perp}^2 = Q^2(1-\frac{Q^2}{s-M_N^2}(\frac{1}{2x_{Bj}}+\frac{M_N^2}{s-M_N^2})).
\ee
One can see that $\vec q _{\perp} \,^2 \le Q^2$ and that it reaches its maximum at $s\gg
Q^2/2x$.
Due to (\ref{kinDIS}), the ISI in DIS can be
interpreted as a smearing of the parton momentum fraction $\xi$
around its parton model value $x_{Bj}$. 
In the following three cases equation (\ref{kinDIS}) simplifies: 
\begin{itemize}
    \item Taking the Bjorken limit:
      \be
      \label{kinBj}
      x_{Bj} = \xi.
      \ee
    \item Neglecting the transverse momentum of the struck parton inside the
    nucleon, but keeping $m^2\ne 0$:
     \be
     x_{Bj} = \xi 
     \frac{Q^2
        }{
          Q^2 - m^2 - 
          \xi (M_N^2 -\frac{m^2}{\xi ^2})\frac{Q^2}{s-M_N^2} }.
     \ee
    \item Taking into account both the parton's transverse momentum and
    off-shellness, but considering the limit $s\gg Q^2/2x$, $s \gg M_N^2$:
     \be
     \label{kinBigS}
     x_{Bj} = \xi 
     \frac{Q^2
        }{
          Q^2 - m^2 +  2 |\vec p_{\perp}| \sqrt{Q^2} \cos (\phi) },
     \ee
     where $\phi$ is the azimuthal angle of the quark momentum. As $Q^2$ goes
     to infinity, equation (\ref{kinBigS}) coincides with (\ref{kinBj}).
\end{itemize}
\item The off-shell partonic cross section is
\be
\label{offshelldis}
\left( \frac{d \hat \sigma }{ d \hat t d \hat u } \right) _{\mbox{off-shell}}=
\frac{2\pi \alpha ^2}{ \hat t^2  \hat s^2} \left( \hat s^2+ \hat u^2 \right) 
\delta (\hat s +\hat u + \hat t - m^2),
\ee
where $\hat u$ and $\hat s$ depend on $m^2$ via
(\ref{off-shell-Mandelstam}) and (\ref{kinDIS}).

\end{itemize}

Therefore, the leading order
expression for the Lorentz invariant DIS cross section (\ref{pDIS}) 
is modified by the ISI as follows: 

\be
\label{offshellDIS}
\left( \frac{d \sigma}{ d t d \hat u } \right) _{\mbox{\small ISI}} \!
= \sum _i e_i ^2 
\int \limits _{0} ^{\infty} \! dm \, \mbox{A} (m,\Gamma) 
\! \int \limits _0 ^1  \!  d\xi
\! \int \!  d\vec p_\perp \,
f_i (\xi , \vec p_\perp ) 
\left(
\frac{d\sigma }{d t d \hat u}
\right) _{\mbox{\small off-shell}}
\ee

To compare to the experiment or to the leading order cross section (\ref{pDIS}), 
we also need to change variables  from partonic $\hat u$ to hadronic $u$ or $x_{Bj}$ ($x_{Bj}$
is related to the hadronic Mandelstam variables ($s$, $t$, $u$) 
by (\ref{bj})).
We choose the following independent variables for the hadronic cross section:
\be
s,\  t,\  x_{Bj}.
\ee
The partonic cross section depends on:
\be
\label{variables}
s,\  t,\  \hat u,\ m^2,\ \xi,\  \vec p_{\perp}.
\ee
We have related the partonic $\hat s$ to $s$ by (\ref{off-shell-Mandelstam}).
The transformation from one set of variables to the other is done in the
following way: 
\be
\label{hatU}
\left( \frac{d \sigma}{ d t d x_{Bj} } \right) _{\mbox{\small ISI}} \! =
\! \int \! d \hat u 
\left( \frac{d \sigma}{ d t d \hat u } \right) _{\mbox{\small ISI}} 
\delta (x_{Bj} - x_{Bj} (s,t,\hat u)),
\ee
where $(d \sigma / d t d \hat u)$ is given by (\ref{offshellDIS}) and 
$x_{Bj}$ as a function of the variables (\ref{variables}) is defined by
(\ref{kinDIS}).
We note in passing that $(d \sigma / dtd x_{Bj})$ is negative, while 
$(d\sigma /dtd\hat u)$ is positive. This has to be taken into account in (\ref{hatU}) by 
an appropriate choice of integration boundaries.
From equations (\ref{offshelldis}), (\ref{offshellDIS}), (\ref{hatU}), we obtain:
\bea
\left( \frac{d \sigma}{ d t d x_{Bj} } \right) _{\mbox{\small ISI}}\! 
= \! \sum _i \frac{ 2 \pi \alpha ^2 e_i ^2 }{t^2}
\! \int \limits _{0} ^{\infty} \!  dm  \, \mbox{A} (m,\Gamma)
\! \int \limits _0 ^1 \! d\xi
\! \int \! d\vec p_\perp \,
f_i (\xi , \vec p_\perp ) 
\! \int d \hat u \hspace{1.5 cm}
\nn
\frac{( \hat s^2 + \hat u^2)}{ \hat s ^2} \, 
\delta \left( \hat s \! + \! \hat u \! + \! t \! - \! m^2  \right) 
\delta \left( x_{Bj}-x_{Bj} (s,t,\hat u,\xi,m^2,\vec p_{\perp}) \right),
\eea
where $\hat s = \xi (s-M_N^2)+m^2$.
The integration over $\hat u$ can be done using one of the
$\delta$-functions. The result is: 
\bea
\left( \frac{d \sigma}{ d t d x_{Bj} } \right) _{\mbox{\small ISI}}\! 
= \! \sum _i \frac{ 2 \pi \alpha e_i ^2 }{t^2}
\! \int \limits _{0} ^{\infty} \!  dm  \, \mbox{A} (m,\Gamma)
\! \int \limits _0 ^1 \! d\xi
\! \int \! d\vec p_\perp \,
f_i (\xi , \vec p_\perp ) \hspace{1.5 cm}
\nn
\times
\left( 1 + \frac{(Q^2-\xi (s-M_N^2))^2}{ (\xi (s-M_N^2)+m^2) ^2} \right)  \,
\delta \left( x_{Bj}-x_{Bj} (s,t,\hat u,\xi,m^2,\vec p_{\perp}) \right),
\eea
where $x_{Bj} (s,t,\hat u,\xi,m^2,\vec p_{\perp})$ is given by (\ref{kinDIS})
and $\hat u =-t-\xi (s-M_N^2)$.
The $\delta$-function can be used to perform the integration over the 
azimuthal angle of the parton momentum. 
The remaining three integrations must be performed numerically. The
limit $s\gg M_N^2, Q^2/2x$ was taken for simplicity. For the unintegrated parton
distributions $f(\xi,\vec p_{\perp})$ we use the factorized form
(\ref{unintPDF}) discussed in more detail in the next section. 
The results for DIS are presented in Sec. \ref{DIS.results}.

%%%%%%%%%%%%%%%%%%%%%%%%%%%%%%%%%%%%%%%%%%%%%%%%%%%%%%%%%%%%%%%%%
\subsection{Drell-Yan process}

\label{m_DY}

We applied 
the same technique to calculate the cross section of the 
Drell-Yan process ($p p \to X + l^+ l^-$). 
In this case, an off-shell quark-antiquark pair
annihilates into a pair of leptons. The virtuality of the quark
(antiquark) coming from the target proton ($m_1^2\equiv p_1^2$) and
that of the antiquark (quark) coming from the projectile proton ($m_2^2 \equiv p_2^2$) 
are in general not equal. 
We assume, however, that their distributions $A(m)$ are the same.

The connection between the observables and partonic variables in case
of two off-shell particles is more complicated. Moreover, the choice of 
proper partonic variables is frame dependent.
We obtain the following kinematic equations in the hadron center of mass 
system:
\bea
\label{kinCMS}
M^2=m_1^2+m_2^2 + \xi_1 \xi_2 P_1 ^- P_2 ^+
+ \frac{\left( m_1^2 +\vec{p}\,^2_{1\perp} \right)
        \left( m_2^2 +\vec{p}\,^2_{2\perp} \right)}{\xi_1\xi_2 P_1^- P_2^+} 
- 2 \vec{p}_{1\perp} \cdot \vec{p}_{2\perp};
\nn
x_F = \frac{\sqrt{S}}{S-M^2}
\left(
\xi_2 P_2 ^+  - \xi_1 P_1 ^-
+ \frac{\left( m_1^2 +\vec{p}\,^2_{1\perp} \right)}{\xi_1 P_1 ^-}
- \frac{\left( m_2^2 +\vec{p}\,^2_{2\perp} \right)}{\xi_2 P_2 ^+}
\right) .
\phantom{M^22=}
\eea
Here, we have used: 
\be
\label{CMS_xi}
\xi_1 = p_1^- / P_1^-, \hspace{1 cm} \xi_2 = p_2^+ / P_2^+,
\ee 
$P_1$ ($P_2$) is the 4-momentum of the target (projectile) hadron,
$p_{1,2}$ denote momenta of the annihilating quark and antiquark, $M^2$ is
the invariant mass squared of the produced leptons. $S$ denotes the hadron 
center of mass energy squared. The Feynman variable is defined as 
$x_F\equiv p_z/(p_z)^{max}$, where $\vec{p}$ is the lepton pair 
momentum. In some works, an approximate definition for the Feynman variable is
used: $x_F\approx 2p_z/\sqrt{S}$. 
We used the exact definition \cite{PDG}
that can be written in the hadron center of 
mass system as follows
\be
x_F\equiv\frac{p_z}{(p_z)_{max}}=\frac{2 p_z \sqrt{S}}{S-M^2}.
\ee
Experimentally observed Drell-Yan pairs have small $M^2$ compared to $S$, so
the difference between the two definitions of $x_F$ are small. However, in the
Drell-Yan scaling limit, $M^2\sim S$ and the difference is finite 
(see formula (\ref{xFlimit}) in Appendix \ref{AppA}).

One sees that the definition of $\xi_2$ is analogous to the DIS case, 
whereas the target's momentum fraction is defined as a ratio 
of minus-components. 
In some articles, alternative definitions are used, for example
$ \xi_1 = p_1^+ / P_1^+, \xi_2 = p_2^+ / P_2^+. $ 
The choice of the definitions (\ref{CMS_xi}) is based on the behavior
of the hadron momenta in the Drell-Yan scaling limit ($S\rightarrow\infty$).
The argument is presented in Appendix \ref{AppA}.
One might prefer to do the calculations in the target rest frame, since
the connection between the observables ($S$, $M^2$, $x_F$, $p_T$) on the one 
hand and the partonic variables ($\xi_i$, $m_i^2$, $p_{i \perp}$) on the 
other is simpler in this case ({\it cf.} (\ref{kinTRF}) in Appendix \ref{AppA}). 
However, factorization in the form (\ref{g_fact}) is not applicable in 
this frame of reference (for a detailed discussion see Appendix \ref{AppA}).

We have calculated the pQCD cross section of the off-shell quark-antiquark 
annihilation into a pair of dileptons:
\bea
\label{DY1}
\frac{ d \hat \sigma}{d\vec{p}\,'_1 d\vec{p}\,'_2} =
\frac{
      e^4 e_q^2  
        \left[
        \hat{t}^2+\hat{u}^2-m_1^4-m_2^4+\hat{s}(m_1+m_2)^2
        \right]
      }{
      16\pi\epsilon '_1 \epsilon '_2 \hat{s}^2 N_c \sqrt{(p_1\cdot p_2)^2-
        m_1 ^2 m_2 ^2}
      }
\delta(p_1+p_2-p'_1-p'_2),
\eea
where $\vec{p}\,'_{1,2}$ are the three-momenta of the leptons, 
$\epsilon \, '_{1,2}$ their energies, and $e_q$ the parton charge in units of the
proton charge, color factor $N_c=3$.

The off-shell partonic cross section, differential over the Drell-Yan
process observables -- mass $M$, Feynman variable $x_F$, and transverse momentum $p_T$
of the lepton pair -- is:
\bea
\label{DY2}
\frac{d \hat \sigma }{dM^2dx_Fdp_T^2} 
=
\! \int \! 
\frac{d\vec{p}\,'_1}{2 \epsilon'_1}\frac{d\vec{p}\,'_2}{2 \epsilon'_2}
d\phi  \
\kappa
\left[ \hat{t}^2+\hat{u}^2-m_1^4-m_2^4+\hat{s}(m_1+m_2)^2 \right]
%\hspace{10.5 cm}
\nn
\times \delta(p_1+p_2-p'_1-p'_2) \delta(p-p'_1-p'_2);
\eea
\be
\label{DY3}
\kappa = 
\frac{\alpha ^2 e_q^2 \left( S- M^2 \right) 
     }{
       \sqrt{S} E M^4 8 N_c \sqrt{(p_1\cdot p_2)^2-m_1^2 m_2^2}
      }.
\ee 
After performing analytically the seven integrations over non-measured 
quantities, four
$\delta$-functions are integrated out and the remaining four 
preserve the correct relation between the hadronic and partonic 
variables ({\it cf.} (\ref{kinCMS})):
\bea
\label{DY4}
\frac{d \hat \sigma }{dM^2dx_Fdp_T^2} =
\hspace{10 cm} 
\nn
\kappa'
\left[
      \frac{M^2}{8} \left( M^2 + \left( m_1+m_2 \right) ^2 \right)
      +\frac{E^2}{6} \left( 4 \, \epsilon _1 ^2 - m_1^2 \right)
      -\frac{E}{3} \, \epsilon _1 \left( M^2 +m_1^2 -m_2^2 \right)
\right] 
\nn
\times \delta \left( M^2 
        -m_1^2-m_2^2 - \xi_1 \xi_2 P_1 ^- P_2 ^+
        - \frac{\left( m_1^2 +\vec{p}_{1\perp}^2 \right)
                \left( m_2^2 +\vec{p}_{2\perp}^2 \right)}{\xi_1\xi_2 P_1^- P_2^+} 
        + 2 \vec{p}_{1\perp} \vec{p}_{2\perp}
\right) 
\nn
\times \delta\left(
x_F - \frac{\sqrt{S}}{S-M^2} \left\{
\xi_2 P_2 ^+  - \xi_1 P_1 ^-
+ \frac{\left( m_1^2 +\vec{p}_{1\perp} ^2 \right)}{\xi_1 P_1 ^-}
- \frac{\left( m_2^2 +\vec{p}_{2\perp} ^2 \right)}{\xi_2 P_2 ^+}
\right\}
\right)
\nn
\times 
\delta \left( \left( \vec{p}_{1\perp}+\vec{p}_{2\perp} \right)^2 -p_T^2 \right) 
\delta \left( E -\epsilon_1 -\epsilon_2 \right);
\phantom{A}
\eea
\be
\label{DY5}
\kappa' = \frac{\alpha^2 e_q^2 E}{M^4 N_c \sqrt{(p_1\cdot p_2)^2-m_1^2 m_2^2}}.
\ee
In (\ref{DY4}), 
$$
\epsilon _1 
\equiv \frac{1}{2} 
\left ( \xi_1 P_1^- + \frac{(m_1^2+\vec p \, ^2_{1 \perp})}{ \xi _1 P_1^-} \right)
\mbox{, } \ 
\epsilon _2 
\equiv \frac{1}{2} 
\left( \xi_2 P_2^+ + \frac{(m_2^2+\vec p \, ^2_{2 \perp})}{ \xi _2
    P_2^+}\right) .
$$

Using the ansatz (\ref{g_fact}) for the case of two
off-shell partons in the initial state, we obtain the hadronic cross section
by integrating over the masses and transverse momenta of quark and antiquark:
\bea
\label{DY6}
\frac{d \sigma }{dM^2dx_Fdp_T^2} =
\sum _i
\! \int \! d \vec{p}_{1\perp} 
\! \int \! d \vec{p}_{2\perp}  
\! \int _0 ^\infty \! d m_1 
\! \int _0 ^\infty \!  d m_2 
\! \int _0 ^1 \!
d \xi_1
\! \int _0 ^1 \! 
d \xi_2 
\mbox{A} (m_1) \mbox{A} (m_2)
\nn
\times
f_i(Q^2,\xi_1,\vec{p}_{1\perp}) \bar f_i(Q^2,\xi_2,\vec{p}_{2\perp})
\frac{d \hat \sigma }{dM^2dx_Fdp_T^2}.\ 
\eea
The integration in (\ref{DY6}) is 8-fold, $d \hat \sigma$ 
is given by equations (\ref{DY4}) and (\ref{DY5}). 
The common parametrization for the unintegrated parton distributions is 
\cite{kt1,kt2,D.definition}
\be
\label{unintPDF}
f (Q^2 , \vec{p}_\perp , \xi )=f(\vec{p}_\perp) \cdot q (Q^2,\xi),
\ee
where
\be
 f(\vec{p}_\perp) = \frac{1}{4 \pi D^2} \exp \{ - \frac{ \vec{p} \,_{\perp}^2}{4 D^2} \} , 
\label{D}
\ee
and $q (Q^2,\xi)$ is the conventional parton distribution. For the latter, we have used the latest 
parametrization by Gl\"uck, Reya, Vogt~\cite{grv}. The mean primordial transverse momentum
of partons is 
\be
<\vec{p}\,_\perp  ^2 > = 4 D ^2.
\label{K_T_D}
\ee

The Gaussian form of $f(\vec{p}_\perp)$ allows the analytical evaluation of the 
integrals over $\vec{p}_{1\perp}$ and $\vec{p}_{2\perp}$. Then, we are left 
with a four-dimensional integral to be done numerically. In the special case 
of a constant spectral function width, one of the integrals over the off-shellness 
can be reduced to a superposition of special functions (incomplete elliptic integrals).

We compare the result of our model, 
in which the partons in the proton have a finite width,
with the experimental data and with the cross sections obtained by
two other methods (LO pQCD and the intrinsic-$k_T$ approach). 

In $k_T$-factorization, the formula 
\be 
\label{fact4}
d \sigma = f ( \xi _1, \vec p_{\perp 1}) f ( \xi _2, \vec p_{\perp 2}) \otimes 
d \hat \sigma ( \xi _1, \xi _2, \vec p_{\perp 1} , \vec p_{\perp 2})
\ee 
is used, 
where $d \hat \sigma$ is the Born cross section for the $q\bar q$ annihilation
into a pair of leptons, $f (\xi , \vec p_\perp)$ is the unintegrated parton 
distribution defined in (\ref{unintegrated.PDF}). 
A proof for the $k_T$-factorization in the Drell-Yan process is
given in the leading twist
in  \cite{Ji.fact,kt.proof}.
In this case, the primordial transverse momenta of the $q$ and $\bar q$ have 
(in general, non-zero) values defined by these distributions in the same way as 
the usual integrated parton
distributions define the light cone fractions of the parton momenta 
($p^+$ for the projectile parton and $p^-$ for the target parton).
In \cite{Ji.fact,kt.proof}, the fourth component of the parton momentum 
($p^-$ for the projectile parton or $p^+$ for the target parton)
is set to
zero due to the following reason. For large hard scales $M$, the projectile
parton momentum is $p_2=(p_2^+,p_2^-,\vec p_{2\perp})\sim M(1,\lambda ^2,\vec
\lambda)$, where $\lambda=m_2/M$. The parameter $\lambda$ is
small for $M>1$~GeV, since the parton off-shellness and transverse
momentum are related to the inverse of the confinement radius and do not scale
with $M$. There exist several parametrizations of unintegrated parton
distributions $f(\xi, \vec p _{\perp})$.

A phenomenological 'intrin\-sic\--$k_T$  approach' has been developed on the
basis of the $k_T$-fac\-to\-ri\-za\-tion theorem. In this model, the
unintegrated distributions are taken in the form (\ref{D}),(\ref{K_T_D}).
An additional difference from \cite{Ji.fact,kt.proof} is that the smaller light cone
component of the parton momentum is put to its on-shell value: $p_2^-=\vec
p_{2\perp}^2 / p_2^+$, which is small, but not zero.
This approach is well described in the literature
\cite{kt1,kt1a,kt2,kt2a} and proves to be very useful
for the calculation of cross sections and asymmetries of different processes. 
It is obtained from (\ref{DY6}) by putting all parton masses to $0$ and
dropping the mass integrations and spectral functions.

In the works \cite{kt2,kt2a}, the Drell-Yan process was studied in the scope
of the intrinsic-$k_T$ approach.
We can obtain the partonic Drell-Yan cross section in the $k_T$-factorization
approach (with on-shell partons) by putting $m_1^2=m_2^2=0$ in (\ref{DY4}), 
(\ref{DY5}). In particular, the following kinematic relations are ensured by
the $\delta$-functions in this case:
\bea
\label{kinKT}
M^2=\xi_1 \xi_2 P_1 ^- P_2 ^+
+ \frac{\vec{p}\,^2_{1\perp} \vec{p}\,^2_{2\perp}
          }{\xi_1\xi_2 P_1^- P_2^+} 
- 2 \vec{p}_{1\perp} \cdot \vec{p}_{2\perp};
\hspace{1cm}
\nn
x_F = \frac{\sqrt{S}}{S-M^2}
\left(
\xi_2 P_2 ^+  - \xi_1 P_1 ^-
+ \frac{\vec{p}\,^2_{1\perp} }{\xi_1 P_1 ^-}
- \frac{\vec{p}\,^2_{2\perp} }{\xi_2 P_2 ^+}
\right) .
\eea
The authors of \cite{kt2,kt2a} 
used the parton model
partonic cross section and the approximate kinematical relations
(\ref{partonMxF}) for simplicity. In section \ref{DY.results}, 
we compare the cross section calculated in our model with off-shell
partons to the result of the $k_T$-factorization approach. 
In order to perform such a comparison, we have calculated 
the Drell-Yan cross section in the intrinsic-$k_T$ approach 
by using the full on-shell partonic cross section and the
exact kinematics (\ref{kinKT}).

%%%%%%%%%%%%%%%%%%%%%%%%%%%%%%%%%%%%%%%%%%%%%%%%%%%%%%%%%%%%%%%%%
\section{Results}
\label{results}

\subsection{DIS}

\label{DIS.results}

\begin{figure}
\begin{center}
\resizebox{0.7\textwidth}{!}{%
  \includegraphics{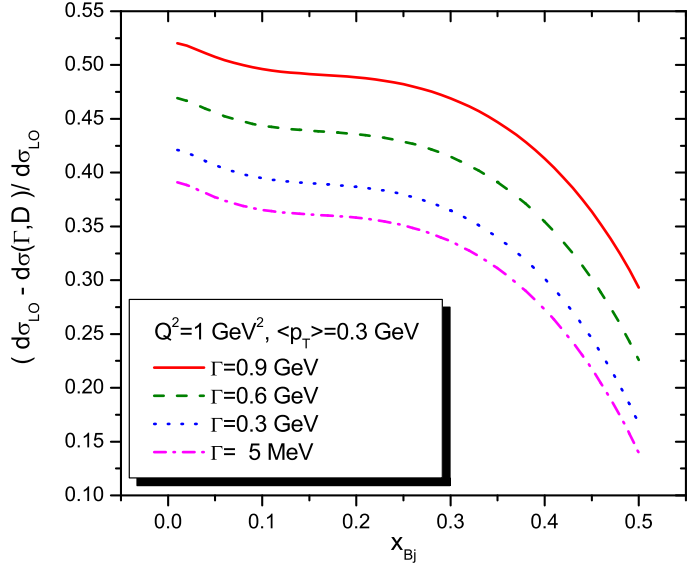}
}
\end{center}
\caption{Relative deviation of the calculated DIS cross section from 
 eq.(\ref{pDIS}) 
for the range of parton spectral function widths $5$~MeV to $0.9$~GeV. 
Mean intrinsic transverse momentum is $0.3$~GeV. $Q^2=1$~GeV$^2$,
$s\gg Q^2$.}
\label{dis1}
\end{figure}

\begin{figure}
\begin{center}
\resizebox{0.65\textwidth}{!}{%
  \includegraphics{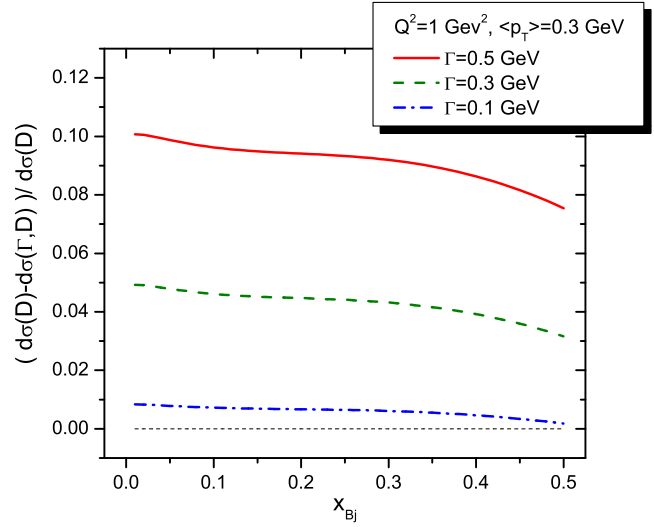}
}
\end{center}
\caption{Relative deviation of the calculated DIS cross section from the 
result of the on-shell calculations taking into account only intrinsic
transverse momentum effects. $Q^2=1$ GeV$^2$, $s\gg Q^2$.}
\label{dis2}
\end{figure}

\begin{figure}
\begin{center}
\resizebox{0.7\textwidth}{!}{
  \includegraphics{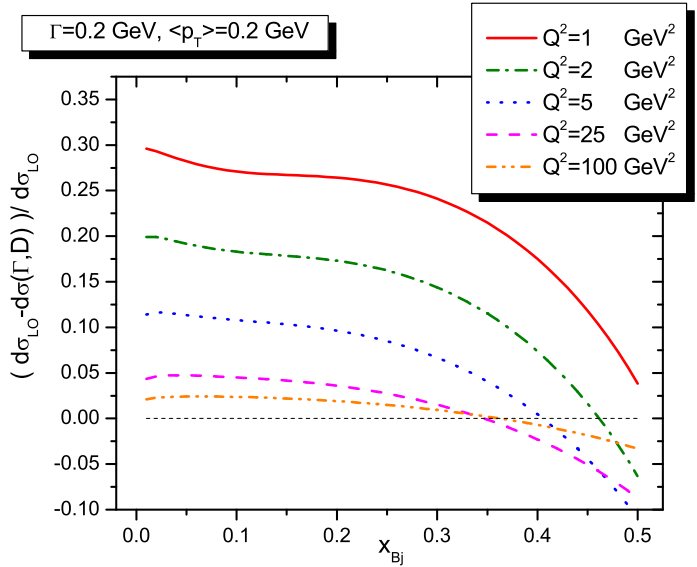}
} 
\end{center}
\caption{Relative deviation of the calculated DIS cross section 
from eq.~(\ref{pDIS}) for
different $Q^2$. The spectral function width is fixed to $\Gamma\! =\! 0.2$~GeV
and the transverse momentum dispersion to $D\! =\! 0.1$~GeV, 
$s\gg \! Q^2$.} 
\label{dis3}
\end{figure}

The results of our calculations for the deep inelastic electron-proton
scattering cross section 
for a range of widths as compared 
to the parton model (eq.~(\ref{pDIS})) are shown in fig.~\ref{dis1}. 
We have found that there is a moderate effect of the initial state interaction
in DIS in the region of small Bjorken $x_{Bj}$ and low momentum transfer $Q^2$. 
The cross section deviation reaches  50\% at 
$Q^2=1$~GeV$^2$, if the parton spectral function width and mean transverse
momentum are both equal to $300$~MeV. 
In fig.~\ref{dis1}, one can also see that the cross section calculated in our
model differs  from the LO even when the parton width is negligibly small
($5$~MeV). This effect is due to the non-vanishing intrinsic transverse momentum.

To separate the effects of the parton off-shellness from those of the intrinsic
transverse momentum, we plot the relative
difference between the result of our model with off-shell partons and the
calculations taking into account only the intrinsic transverse motion 
(fig.~\ref{dis2}). To
obtain the cross section in the latter approach, we put $\Gamma$ to zero in 
the formulas of section~\ref{m_DIS} thus forcing the parton on-shell.
It is seen that this difference amounts to at most 10~\% of the cross section.

For values of $Q^2$ above 25~GeV$^2$,  
the initial state interaction in DIS gives
at most a 5\% deviation from the lowest order cross section (\ref{pDIS}).
The $Q^2$-suppression of parton virtuality and intrinsic transverse 
momentum effects
in DIS is illustrated in
fig.~\ref{dis3}. For most of the experimentally investigated 
values of $Q^2$, the ambiguity in the parton distribution function 
parameterizations due to the renormalization scale uncertainty is of 
the same order as the ISI effect in DIS. 

The difference between the off-shell result and the leading order
cross section at $Q^2\ge2$~GeV$^2$ is too small to be resolved by the present 
experiments. 
In the region of $Q^2\le2$~GeV$^2$, the difference is $30-40$~\%, which should
be observable. 
However, in order to make a quantitative comparison to the experiment 
at such low $Q^2$ and $x_{Bj}$, we would have to incorporate into our model 
other effects, such as resonance production and diffractive scattering
\cite{CLAS,duality,diffractive}. 
We conclude that, using the model described 
in the present paper, we cannot extract the value of the parton width in the nucleon
from the DIS data. This is the result expected by the
analogy to nuclear physics, because the DIS cross section is too
inclusive. On the other hand, the DIS data do not contradict the 
assumption of a finite parton width in the proton.

%%%%%%%%%%%%%%%%%%%%%%%%%%%%%%%%%%%%%%%%%%%%%%%%%%%%%%%%%%%%%%
\subsection{Drell-Yan process}

\label{DY.results}

In contrast to the DIS case, the effect of parton off-shellness on the transverse 
momentum distribution of the Drell-Yan lepton pairs is substantial.
In this section, we present the Drell-Yan triple differential cross
section calculated by the method described in Section \ref{m_DY}. We
compare the result of our model, in which the partons in the proton have a finite width,
with the experimental data and with the cross sections obtained in
two other approaches (LO pQCD and standard $k_T$-factorization). 

Calculations using LO pQCD and collinear factorization analogous to (\ref{fact})
\be 
\label{fact3}
d \sigma = f ( \xi _1) f ( \xi _2) \otimes 
d \hat \sigma ( \xi _1, \xi _2)
\ee
give a simple result for the triple differential Drell-Yan cross section ($p_T$-distribution of the
dileptons)  - a $\delta$-function at zero $p_T$. This follows from
the fact that the annihilating $q$ and $\bar q$ in this approach are collinear
with the corresponding hadrons, thus the $q\bar q$ pair has no transverse
momentum in the hadron center of mass system. 
%Since both $q$, $\bar q$ are on-shell, 
Therefore, 
the resulting lepton pair cannot gain any transverse momentum in
this model. In contrast, the experimentally measured transverse momentum 
distribution of the dileptons extends to $p_T=4$~GeV at a hard scale (the
mass of the lepton pair $M$) as high as $8.7$~GeV. Note that NLO corrections 
do not cure the disagreement with the data. 
The Drell-Yan pair $p_T$-distribution obtained in fixed order pQCD is
divergent at $p_T=0$.
A resummation of an infinite series of diagrams is necessary to 
obtain a finite value for the triple differential Drell-Yan cross section at
$p_T=0$ in pQCD with on-shell partons~\cite{logM/P}. 
The resummed pQCD cross section is in qualitative agreement with 
experiment~\cite{logM/P}.

In order to analyze the effect of a finite parton width and distinguish it
from the effect of the intrinsic transverse momentum, we have performed the 
calculations in both the intrinsic-$k_T$ approach and in our model allowing
for off-shell partons. 
We used the formalism developed in Sec.~\ref{m_DY} to 
calculate the cross section of 
the Drell-Yan process in the kinematics of the experiment E866 \cite{exp0,exp}
in the intrinsic-$k_T$ approach and 
for a parton off-shellness
distributed according to the Breit-Wigner spectral function (\ref{BW}).

\begin{figure}
\begin{center}
\resizebox{0.65\textwidth}{!}{%
  \includegraphics{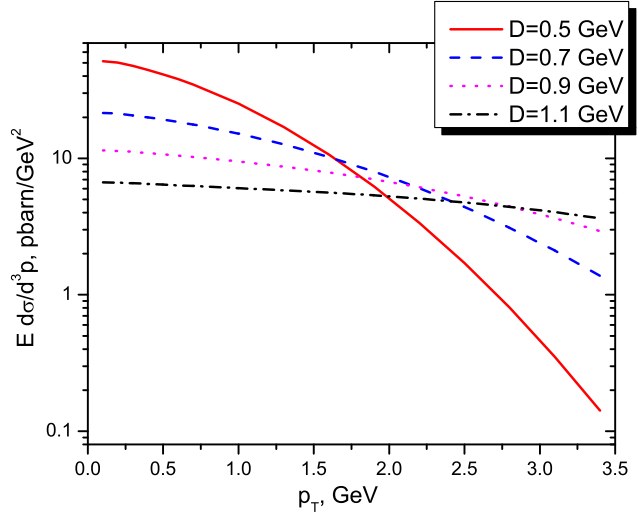}
} 
\end{center}
\caption{Calculated distribution of the Drell-Yan lepton pair's
transverse momentum in the $k_T$-factorization approach
for different values of the parton primordial
transverse momentum dispersion. $M=4$~GeV, $x_F=0.1$, $\Gamma=0$.}
\label{var2}
\end{figure}

\begin{figure}
\begin{center}
\resizebox{0.65\textwidth}{!}{%
  \includegraphics{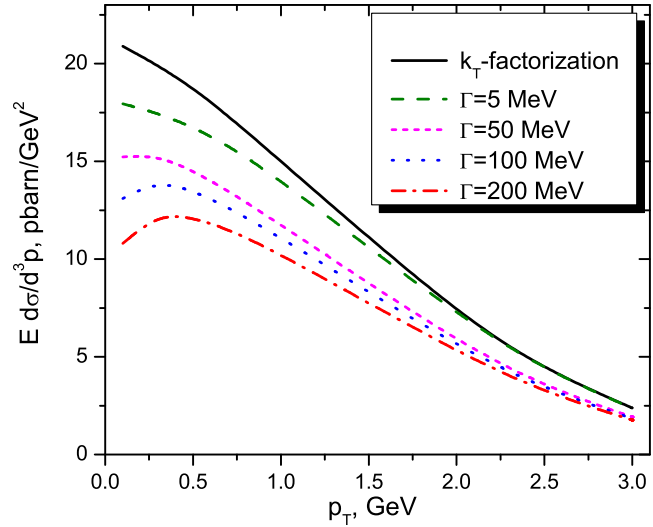}
} 
\end{center}
\caption{Calculated distribution of the Drell-Yan lepton pair's
transverse momentum in our model for different values of 
the parton spectral function width. $M=4$~GeV,
$x_F=0.1$, $D=0.7$. The solid line gives the result of a calculation with $\Gamma=0$.}
\label{var1}
\end{figure}

We present the obtained cross sections for different values of the parameters in
figures \ref{var2} and \ref{var1}. 
We illustrate in fig.~\ref{var2} that the slope of the distribution mainly 
depends on the dispersion of the intrinsic transverse momentum ($D$), 
which is proportional
to the primordial transverse momentum of the parton (see (\ref{K_T_D})). 
In the limit, in which the dispersion of the intrinsic transverse momentum 
($D$) goes to zero,
the leading order result of perturbative QCD, {\it i.e.} a sharp
peak at $p_T=0$, is restored.

On the other hand, the parton width variation leads to
changes of the cross section magnitude and influences the behavior
of the distribution in the region of low $p_T$ (see fig.~\ref{var1}). 
One can also see in figure \ref{var1} that  our model
approaches the result of the standard intrinsic-$k_T$ method
as the parton width ($\Gamma$) goes  to zero. 
At finite width,
the  shape of the cross section obtained in our model 
is different from the result of the
intrinsic-$k_T$ approach in the low $p_T$ region.
Also, 
the magnitude of the cross section is different.
This indicates that some additional 
nonperturbative effects are included via a finite parton width.

%=============================================fit KT 1:

\begin{figure*}
\begin{center}
\subfigure[$4.2\!\le\! M\!\le\!5.2$~GeV] % caption for subfigure a
{
        \resizebox{0.7\textwidth}{!}{%
        \includegraphics{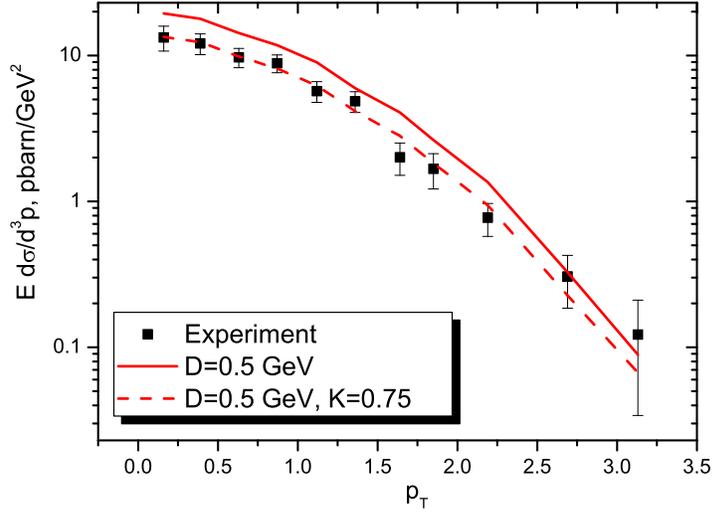}
        } 
        \label{fitKT1.a}
}
\subfigure[$5.2\!\le\! M\!\le\!6.2$~GeV] % caption for subfigure b
{
        \label{fitKT1.b}
        \resizebox{0.7\textwidth}{!}{%
        \includegraphics{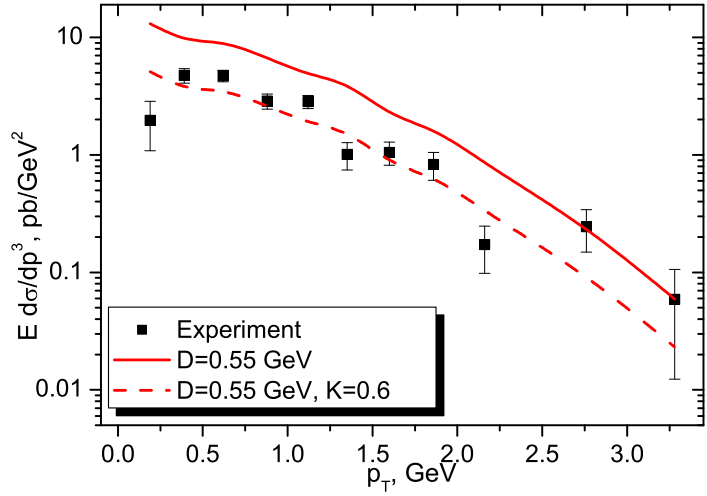}
        } 
}
\end{center}
\caption{Drell-Yan cross section in $k_T$-factorization approach
compared to the data for continuum dimuon production in $pp$
collision from \protect{ \cite{exp}}. Varying the dispersion of intrinsic transverse
momentum (D), the slope of the distribution is fitted (solid line). 
An additional overall K-factor is necessary to reproduce 
the cross section amplitude (dashed line), $ -0.05\!\le\! x_F\!\le\!0.15$.}
\label{fitKT1} 
\label{fit.begin}
\end{figure*}

\begin{figure*}
\begin{center}
\subfigure[$6.2\!\le\! M\!\le\!7.2$~GeV] % caption for subfigure c
{
        \resizebox{0.7\textwidth}{!}{%
        \includegraphics{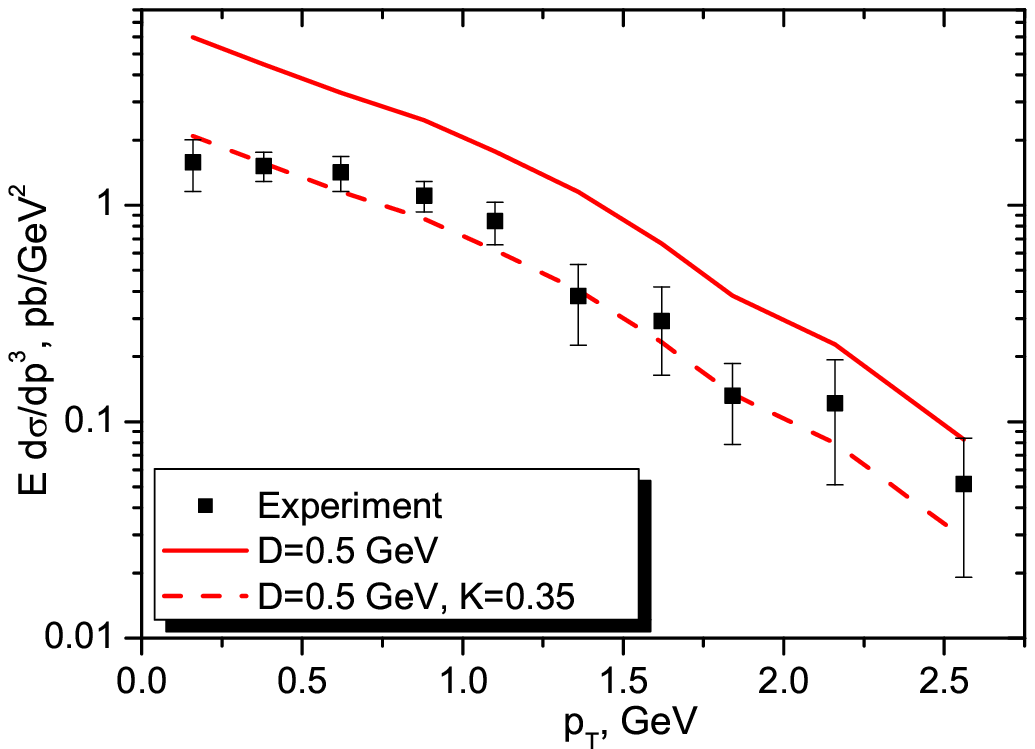}
        } 
        \label{fitKT1.c}
}
\subfigure[$7.2\!\le\! M\!\le\!8.7$~GeV] % caption for subfigure d
{
        \label{fitKT1.d}
        \resizebox{0.7\textwidth}{!}{%
        \includegraphics{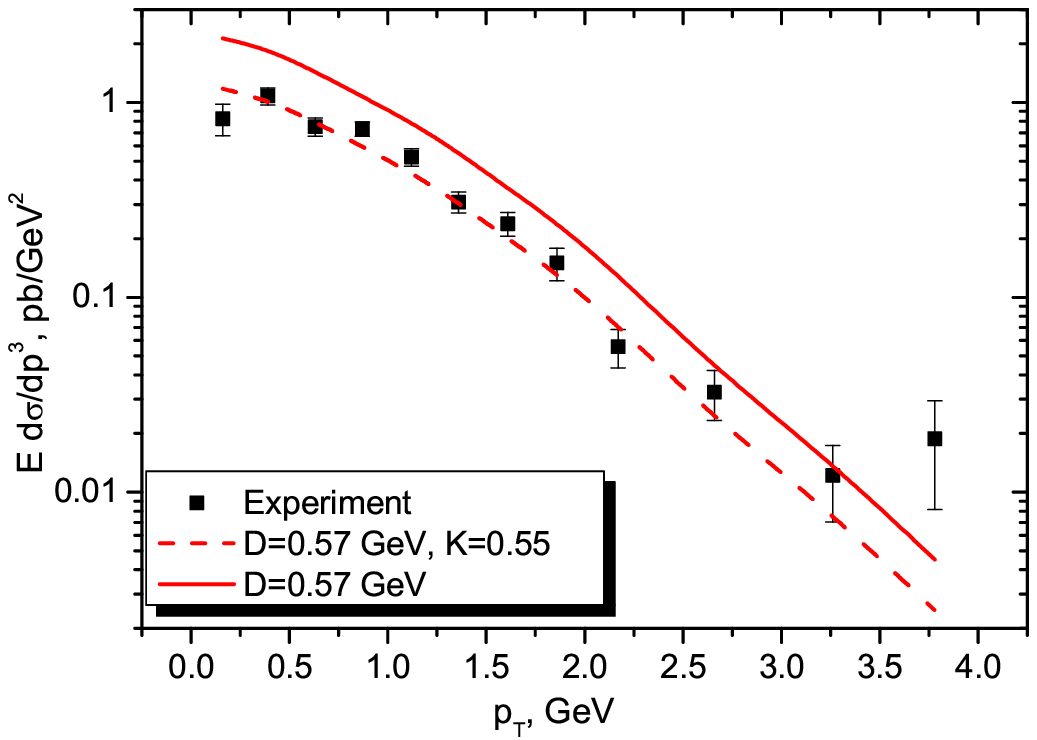}
        } 
}
\end{center}
\caption{Same as Fig. \ref{fitKT1}, but higher mass bins.}
\label{fitKT2} 
\end{figure*}

\begin{figure}
\begin{center}
        \resizebox{0.7\textwidth}{!}{%
        \includegraphics{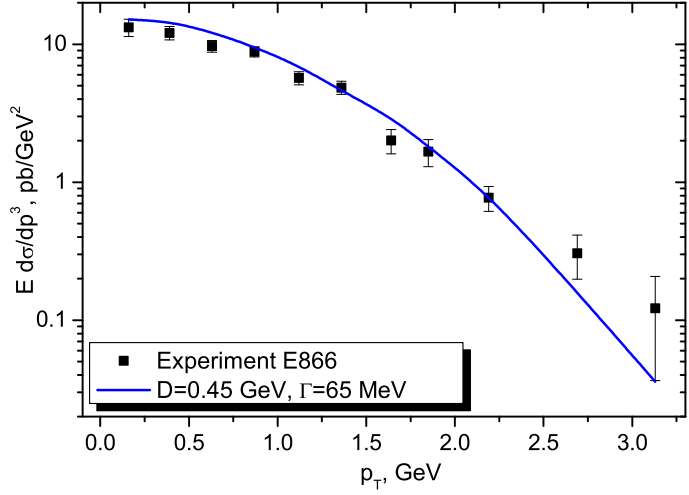}
        } 
\caption{The Drell-Yan cross section as calculated in our model (solid line) 
compared to the data of 
the Fermilab experiment E866 for the continuum dimuon production in 800~GeV proton, 
$4.2\!\le\! M\!\le\!5.2$~GeV, $-0.05\!\le\! x_F\!\le\!0.15$.}
\label{fit1} 
\end{center}
\end{figure}

\begin{figure}
\begin{center}
        \resizebox{0.7\textwidth}{!}{%
        \includegraphics{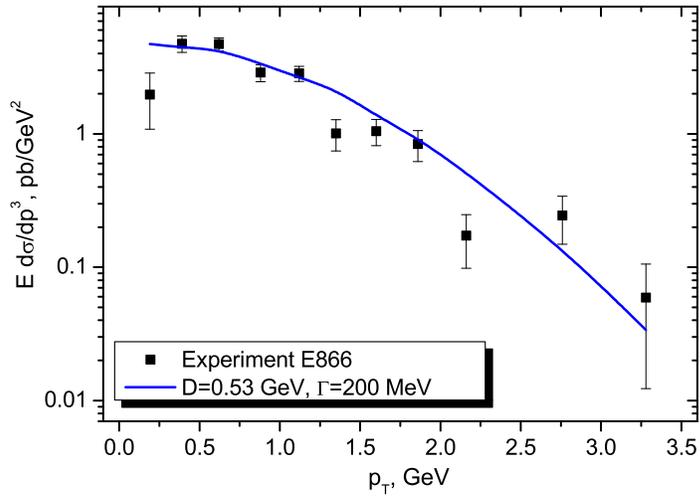}
        } 
\end{center}
\caption{Same as fig.~\ref{fit1}, only for a higher mass bin: 
$5.2\!\le\! M\!\le\!6.2$~GeV.}
\label{fit2} 
\end{figure}

\begin{figure}
\begin{center}
        \resizebox{0.7\textwidth}{!}{%
        \includegraphics{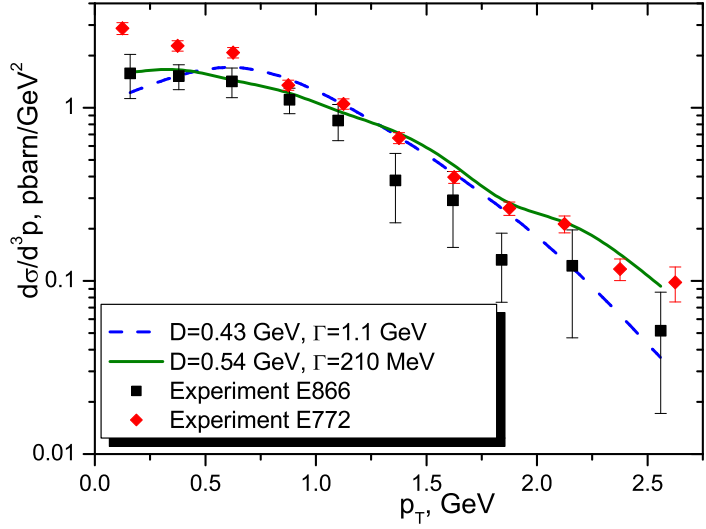}
        } 
\end{center}
\caption{
The Drell-Yan cross section as calculated in our model 
compared to the data of E866 on $pp\! \to\! \mu^+\mu^-X$,
$\ 6.2\!\le\! M\!\le\!7.2$~GeV, $-0.05\!\le\! x_F\!\le\!0.15$, 
and to the  data of E772 on $pd\! \to\! \mu^+\mu^-X$, 
$\ 6\!\le\! M\!\le\!7$~GeV, $0\!\le\! x_F\!\le\!0.3$. See main text for
more details about the different lines.}
\label{fit3} 
\end{figure}

\begin{figure}
\begin{center}
        \resizebox{0.7\textwidth}{!}{%
        \includegraphics{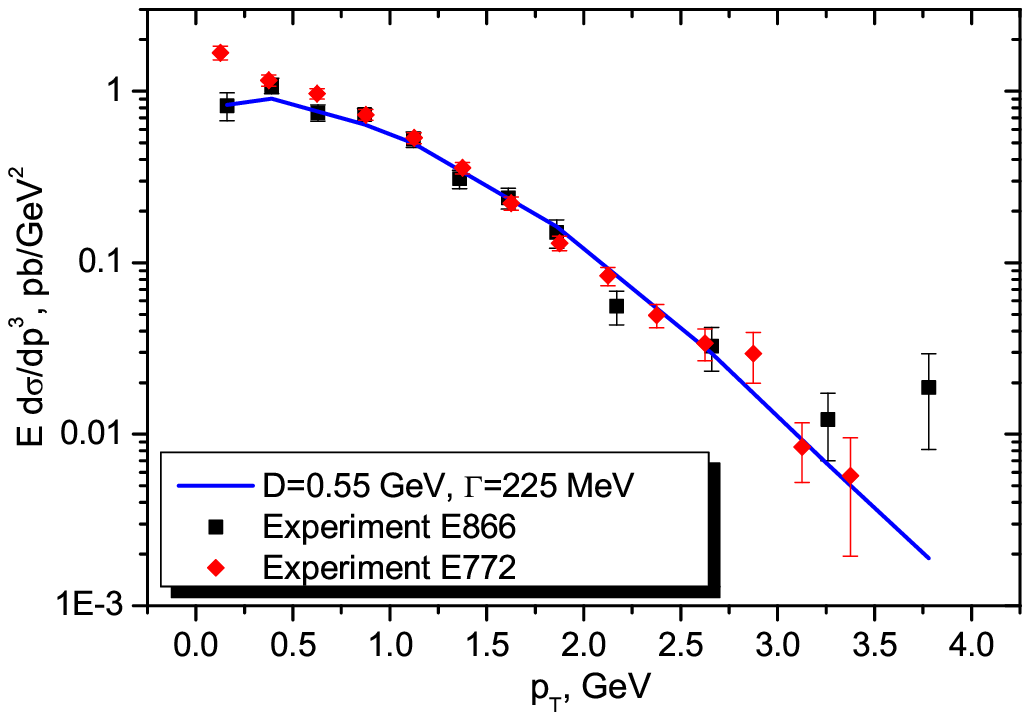}
        } 
\end{center}
\caption{Same as fig.~\ref{fit3}, only for a higher mass bin: 
$7.2\!\le\! M\!\le\! 8.7$~GeV.}
\label{fit4} 
\label{fit.end}
\end{figure}

In figures~\ref{fit.begin}-\ref{fit.end}, calculations both in the
model with off-shell partons and in the standard 
$k_T$-fac\-to\-ri\-za\-tion approach are compared to the data
of the Fermilab experiment E866 for the continuum dimuon production in $pp$
collisions at $800$~GeV incident energy. In this experiment, 
both the double differential Drell-Yan cross section 
 $d\sigma /dM^2 dx_F$ (data published in \cite{exp0}) 
and the triple differential cross section $d\sigma /d\vec p$ 
(data published in \cite{exp}) were
measured in a wide range of $M$ and $x_F$ ($\vec p$ is the lepton
pair's momentum). The $p_T$-distribution was obtained in terms of 
the triple differential cross section averaged over the azimuthal angle of the lepton pair
\be
 \frac{d\sigma}{d\vec p}
\equiv \frac{2}{\pi \sqrt{s}}\frac{d\sigma}{d x_F d p_T ^2}
=\frac{2}{\pi \sqrt{s}} \int \limits _{\mbox{\small bin}} \! 
                                \frac{d\sigma}{d x_F d p_T ^2 dM^2} \, dM^2 .
\label{triple}
\ee
The data points were averaged in several bins in  $M$ and $x_F$.
The $x_F$ binning is responsible for the wiggly structures both in the data
and some of our calculations.

%=============================================fit 1:

%=============================================fit 2:

%=============================================fit 3:

%=============================================fit 4:

%=============================================

The result of the standard $k_T$-factorization approach is shown in
figures~\ref{fitKT1} and~\ref{fitKT2} (solid line). The slope of the 
cross section can be reproduced by an appropriate choice of the single
parameter ($D$) of the
intrinsic transverse momentum distribution, given by (\ref{D}). 
The optimal values for $D$ are $500-600$~MeV, which correspond to 
\be
<p_\perp ^2>^{1/2} = 1.0-1.2 \mbox{ GeV}.
\ee
A slightly smaller value for this parameter was obtained in \cite{kt2,kt2a}
from the analysis of the data of the experiment E744 on Drell-Yan
cross section in $pp$~collisions at $400$~GeV incident energy:
\be
<p_\perp ^2>^{1/2} = 0.8-1.0 \mbox{ GeV}.
\label{mean.KT}
\ee

Still, the data are overestimated by a factor of $2-3$, depending on the mass
of the Drell-Yan pair ($M$). Dashed lines in figures \ref{fitKT1} and \ref{fitKT2}
illustrate that the data can be fitted by 
introducing an additional overall factor (K). 
The discrepancy between the calculations and the data is larger for higher $M$. 
Thus, in the $k_T$-factorization approach, the magnitude of the cross section 
cannot be correctly reproduced. An additional overall $K$-factor is
necessary that reflects the importance of higher order corrections to the Drell-Yan
cross section.

In contrast, the calculations with a finite parton width  yield not only the
experimentally measured shape of the cross section but also its
amplitude without any $K$-factor.
Instead, we take care of the higher-twist and NLO effects assumed to be contained in
$K$ by introducing the physically transparent off-shellness ($\Gamma$).
Note that we work in leading order of perturbative QCD concerning the processes that
enter the calculation of the Drell-Yan cross section. 
At NLO of perturbative QCD, additional processes contribute, 
namely, gluon bremsstrahlung ($q\bar q\to l^+l^- g$), gluon
Compton scattering ($gq\to l^+l^- q$ and $g \bar q \to l^+l^- \bar q$), 
and vertex corrections.
The bremsstrahlung, along with diagrams of even higher order and with gluon
exchanges between the active (anti-)quark and the spectators, contributes to
the quark width. Therefore, going to higher orders of standard perturbation
theory, while dressing quark lines with spectral functions, would be a double
counting. In other words, we have already included part of the NLO processes
by our finite width calculation. Some processes beyond the NLO are also 
included.
On the other hand, our model can be improved by taking into account
also gluon Compton scattering and vertex corrections. In this case, the gluon
line should be dressed, too, and the number of model parameters increases. In 
the present paper, we concentrate on the effect of a finite quark width on
observables and consider only the leading reaction mechanisms, {\it i.e.}
$q\bar q$ annihilation for the Drell-Yan process.

The comparison of our results with the data is given in figures
\ref{fit1}-\ref{fit4}.
The values for the average parton primordial transverse momentum 
\be
<p_\perp ^2>^{1/2} = 0.9-1.1 \mbox{ GeV}
\ee
are compatible with those existing in the literature
(\ref{mean.KT}).
Allowing for off-shell partons, we eliminate the need for any
K-factor.
Choosing $\Gamma$ in the order of $100$~MeV ({\it cf.} table \ref{G_vs_M}
for details), both the amplitude and the slope of the cross section 
are well reproduced. 

The dependence of the optimal values for the parameters 
(dispersion $D$ and width $\Gamma$)
on the mass of the Drell-Yan pair was obtained by fitting experimental
data within different bins of $M$ independently. The result is presented in
figures \ref{fit.begin}-\ref{fit.end} and in table \ref{G_vs_M}. 
Note that the varying quality of the data in different mass bins leads to
large uncertainties in the extraction of the width. 
In table \ref{G_vs_M} we present the average values and 
 uncertainties for $D$ and $\Gamma$. The latter have been obtained by
 analyzing the $\chi ^2$ values as a function of $D$ and $\Gamma$.
We find that the optimal $\Gamma$ increases with the
hard scale (the mass of the Drell-Yan pair).
The dependence of $\Gamma$ on $M$ indicates that, at higher scales,
partons with broader spectral functions are probed.
We did not study the dependence of our parameters on $x_F$.

The analysis of the data in the mass bin $6.2\le M\le7.2$~GeV calls for more
discussion. As shown in figure 11, the best fit (dashed line)
 to this data set leads to
values for both parameters (D=0.43~GeV, $\Gamma$=1.1~GeV), which are not in 
 the trend set by the fits to the other three data bins (cf. table~\ref{G_vs_M}).
Thus, we did not trust this fit and sought for more experimental
input. For $p_T\ge 1$~GeV, the data of the experiments 
E866 on $pp\to\mu^+\mu^-X$ and E772 \cite{E772} on $pd\to\mu^+\mu^-X$ agree
very well in all the mass bins, except the one of figure~\ref{fit3} 
(see figure~\ref{fit4}, for example). 
Therefore, we compared our fit (dashed line) to the experimental data on the $pd$
cross section from E772 in approximately the same mass range (figure~\ref{fit3}). 
One can see that the calculations with $D=0.43$~GeV and
$\Gamma=1.1$~GeV (dashed line) do not reproduce the high-$p_T$ part of the $pd$ cross
section. On the other hand, if the trend-average values from
table~\ref{G_vs_M} are applied ($D=0.54$~GeV, $\Gamma=210$~MeV, solid line in
figure~\ref{fit3}), the
cross section calculated in our model both describes the E866 data on the
border of experimental error bars and reproduces the $pd$ cross section of
E772 at $p_T\ge1$~GeV.

%%%-----------------------------------------------------------------------
%this table is in LaTeX 2.e style and requires packages "array"&"hhline"
%%%-----------------------------------------------------------------------
\begin{table*}
\begin{center}
\begin{tabular}{|c||c|c|c|c|}     
\hhline{|=:t:====|}
$M$ & $4.2$-$5.2$ & $5.2$-$6.2$ & $6.2$-$7.2$ & $7.2$-$8.7$ 
%& $10.85$-$12.85$ & $12.85$-16.85   %%%%%%%%%%% - esche stolbcy 
\\ 
\hhline{|=||====|}
$D$        & $450\pm100$ &  $530\pm70$ & 540*    & $550\pm60$   \\  
\hhline{|-||-|-|-|-|}
$\Gamma$   &  $65\pm20$  &  $200\pm75$ & 210*   & $225\pm75$   \\ 
%\hhline{|-||-|-|-|} 
%$\chi^2$   &    &     &     &     &     \\
\hhline{|=:b:====|}
\end{tabular}
\end{center}
\caption{Optimal parameters for different masses of the Drell-Yan pair,
\mbox{$-0.05\!\le\! x_F\!\le\!0.15$}. All values are in MeV. Values denoted
with stars are trend-average and not best-fit. See main text for details.}
\label{G_vs_M}
\end{table*}

%%%----------------------------------------------------------------------
%the following table is in LaTeX 2.09 style
%NB: correlate \rule-argument with \extrarowheight (see above)
%%%----------------------------------------------------------------------
%\begin{table*}
%\begin{center}
%\renewcommand{\arraystretch}{0}%
%\begin{tabular}{|c||c@{\strut\hspace{\tabcolsep}}|c|c|}     
%\hline 
%\multicolumn{4}{|c|}{\rule{0pt}{2pt}}\\
%\hline 
%&\multicolumn{1}{c|}{\rule{0pt}{3pt}}&&\\
%$x_F$     & $-0.05$-$0.15$ & $0.15$-$0.35$ & $0.35$-$0.55$ 
%%& $0.55$-$0.8$ %%%%%%%%%% - esche stolbec
%\\[3pt] 
%\hline
%\multicolumn{4}{|c|}{\rule{0pt}{2pt}}\\
%\hline 
%&\multicolumn{1}{c|}{\rule{0pt}{3pt}}&&\\
%$D$       &     0.45       &               &              \\[3pt] 
%\hline
%&\multicolumn{1}{c|}{\rule{0pt}{3pt}}&&\\
%$\Gamma$  &     0.15       &               &              \\[3pt]
%\hline
%\multicolumn{4}{|c|}{\rule{0pt}{2pt}}\\
%\hline 
%\end{tabular}
%\end{center}
%\caption{Optimal parameters as a function of the Feynman variable $x_F$, 
%\mbox{$4.2\!\le\! M\!\le\!5.2$}. All values are in GeV.}
%\label{G_vs_xF}
%\end{table*}

Allowing for a finite parton width and using a single-parameter form
for the parton spectral function, we account for
non-perturbative effects, including the K-factor. 
The result of the collinear factorization and fixed order pQCD ($\delta$-peak at
$p_T=0$) is not reached in the experiment even at masses of lepton 
pairs as high as $M\sim16$~GeV. 
There is one area of hard scales, where the  intrinsic $k_T$ approach 
seems to reproduce the cross section with good accuracy: at low $M$ the
$K$-factor of the intrinsic-$k_T$ approach 
is closer to $1$. As the dilepton mass goes higher,
the measured distribution is getting more sharply peaked. This suggests that
the result of LO pQCD might be recovered at Drell-Yan pair masses, which are 
higher than those yet observed. On the other hand, our model allowing for
 off-shell partons with finite width works well for all hard scales $M$.

%%%%%%%%%%%%%%%%%%%%%%%%%%%%%%%%%%%%%%%%%%%%%%%%%%%%%%%%%%%%%%%
\section{Summary and outlook}

\label{summary}

The research presented here reveals the importance of the parton initial state
interaction for the analysis of high energy processes. 
We have developed a formalism to study the quark structure of 
hadrons going further than the widely studied picture of collinear 
noninteracting partons. The parton off-shellness effects
missed in the standard treatment were taken into
account by dressing the parton lines with spectral functions and 
using the factorization assumption.
In this way, higher twist corrections to standard pQCD were modelled.

We have calculated the cross sections of deep inelastic $ep$ scattering 
and the Drell-Yan process $pp\to l^+l^-X$ in the model allowing for a finite
parton width. Off-shellness effects arise from the fact that the time-like lightcone
momentum of the parton ($p^-$) is not fixed by an on-shell condition ($p^-=p_\perp^2/p^+$)
or by a collinearity condition ($p^-=0$). Since the partons in the proton
interact, $p^-$ is in fact distributed with some finite width. To disentangle the
off-shellness effects from the effect of the parton primordial transverse
momentum, we have additionally calculated the Drell-Yan cross section in
the standard intrinsic-$k_T$ approach. The obtained cross sections in both
models were compared to the data on the triple differential cross section of
the process $pp\to l^+l^-X$.

We have found a moderate effect of the initial state interaction
in DIS in the region of small Bjorken $x_{Bj}$ and low momentum transfer $Q^2$. 
For a parton width of 300~MeV, the cross section increases due to the
finite quark width in the proton reaches 10\% at 
$Q^2=1$~GeV$^2$. On the other hand, the effect is $Q^2$-suppressed.
For values of $Q^2$ above 10~GeV, the initial parton off-shellness 
generates only at most 2\% of the cross section.
For most of the experimentally investigated values of $Q^2$,
the difference between the off-shell result and the leading order
cross section is too small to be resolved by present experiments. 
We conclude that the value of the parton width in the nucleon cannot 
be extracted from the DIS data, because the DIS cross section is too
inclusive. This is the result expected by the analogy to nuclear physics. 
On the other hand, the DIS data do not contradict the assumption of 
the finite parton width in the proton.

In contrast, we discover a substantial contribution of the parton 
off-shellness to the transverse momentum distribution of the high-mass 
virtual photons produced in high-energy hadron-hadron collisions in the whole
range of hard scales, at which the cross section has been measured. 
The triple differential Drell-Yan cross section is a more
exclusive observable than the DIS cross section. That is why the effect of the
parton off-shell\-ness was expected to be larger in the Drell-Yan
case. Our results confirm this expectation.

Although the intrinsic-$k_T$ approach alone can reproduce the slope of the experimentally
measured distribution of dileptons, an overall K-factor is necessary to fit
the data. 
Shape {\em and magnitude} of the cross section are much better reproduced by a model
that allows for off-shell partons. In particular, one can fit the data without 
a K factor.
The parton width in the proton was estimated from the comparison to the 
data. For a mass of the Drell-Yan 
pair of $4.2-8.7$~GeV, the best fits were obtained with quark (antiquark) width
of $50-250$~MeV and intrinsic transverse partonic momentum dispersion
of $400-600$~MeV. This corresponds to a mean primordial transverse momentum
of the parton inside a proton of $<k_T>=0.8-1.2$~GeV.

Since the Drell-Yan process is expected to be one of the leading background
contributions at the future high energy facilities, it is
important to predict its cross section as precisely as possible.
The obtained triple differential cross section of the dilepton production in 
$pp$ collisions is also a necessary input for models, studying nuclear 
medium via high energy dileptons, produced in $pA$ and $AA$ collisions.
To meet this demand and to consistently evaluate the ISI effects in high energy
processes, we need to improve our knowledge of the parton spectral function in
the nucleon. The single-parameter Breit-Wigner parametrization might be
insufficient. To pin down the parton spectral function, the study of other 
exclusive processes will be necessary. In particular, it should be possible to
reduce the sizable uncertainty in the width.
We shall address this issue in the future.

\section{Acknowledgements}

We gratefully acknowledge helpful discussions with P.~Hoyer and A.~Vainshtein.

%%%%%%%%%%%%%%%%%%%%%%%%%%  Appendix %%%%%%%%%%%%%%%%%%%%%%%%%%%%%%%%%%%
\begin{appendix}
\section{Appendix}

\label{AppA}

In this appendix we present the details of the Drell-Yan process description
in different frames of reference. 
The arguments for the definitions of the partonic momentum fractions
used in our calculations are given as well.

Let us consider the Drell-Yan scaling limit ($S \rightarrow \infty$).
The light cone components of hadron momenta in the center of mass system are
\be
\label{P_in_CMS}
\left( P_1^\mp \right)^2 = \left( P_2^\pm \right)^2
=
\frac{S}{2}-M_N^2\pm \sqrt{ \left( \frac{S}{2} \right)^2-M_N^2S }.
\ee
Thus, the plus-\-com\-po\-nent of the projectile's momentum $P_2^+$ and 
the minus-\-com\-po\-nent of the target's momentum $P_1^-$ go to infinity 
$\sim \sqrt{S}$, while all the other components are negligible in the 
scaling limit. 

With the chosen definitions of $\xi_i$, we get as a limit of (\ref{kinCMS}):
\bea
\label{xFlimit}
M^2=\xi_1 \xi_2 S; \hspace{0.5cm}
\nn
x_F = \frac{\xi_2 -\xi_1}{1-\xi_1 \xi_2 }.
\eea
Applying approximate definition $x_F\approx 2p_z/\sqrt{S}$, we recover
 the well known parton model relations:
\bea
\label{partonMxF}
M^2=\xi_1 \xi_2 S;
\nn
x_F =\xi_2 -\xi_1.
\eea
This means  that we can use $\xi_1 = p_1^+ / P_1^+$, 
$\xi_2 = p_2^- / P_2^-$ as the arguments of the parton distribution
functions in the factorization formula for the Drell-Yan process
in the center of mass system. 

In contrast, hadron light cone momenta scale differently in 
the target rest frame:
\be
\label{P_in_TRF}
P_2^\pm=\frac{S}{2M_N} - M_N \pm \sqrt{\left(\frac{S}{2M_N}\right)^2 -S};
\hspace{1 cm}
P_1^\pm=M_N.
\ee
The motion of the projectile is again confined to the light cone. However, 
there is no special direction for the target parton. Therefore, one needs 
to parameterize the "soft" properties of the target with more general 
distribution functions, which depend on the 4-momentum of the parton 
instead of a single scalar variable $\xi$. These functions $W(p)$ (partonic
Wigner distributions) were introduced by
X.~Ji in~\cite{ji}.

M.~Sawicki and J.P.~Vary \cite{error} considered the Drell-Yan process 
in the target rest frame within the factorization framework. They used 
the analogous to DIS definitions of the both momentum fractions,  
\be
\label{vary_xi}
\xi_1= p_1^+ / P_1^+; \hspace{1 cm} \xi_2= p_2^+ / P_2^+;
\ee
and found scaling violation. 
Indeed, in the target rest frame (\ref{kinCMS}) transforms to 
\bea
\label{kinTRF}
M^2=m_1^2+m_2^2 
+ \frac{\xi_1 M_N}{\xi_2 P_2^+} \left( m_2^2 +\vec{p}_{2\perp}^2 \right)
+ \frac{\xi_2 P_2^+}{\xi_1 M_N} \left( m_1^2 +\vec{p}_{1\perp}^2 \right)
- 2 \vec{p}_{1\perp} \cdot \vec{p}_{2\perp};
\nn
x_F = \frac{1}{\omega}
\left(
\xi_1 M_N  + \xi_2 P_2 ^+
- \frac{\left( m_1^2 +\vec{p}_{1\perp}^2 \right)}{\xi_1 M_N}
- \frac{\left( m_2^2 +\vec{p}_{2\perp}^2 \right)}{\xi_2 P_2 ^+}
\right) ,
\phantom{M^2=m_1^2+m_2^2 }
\eea
where we used the definitions (\ref{vary_xi}) and
\be
\omega=\frac{\sqrt{S}}{2} 
\left(
\frac{S}{2M_N^2} - 1 + \frac{M^2}{S}
\right)
/ \sqrt{\frac{S}{4M_N^2} -1}.
\ee
Bearing in mind equations (\ref{P_in_TRF}) and (\ref{kinTRF}), we arrive 
at the following limiting values for $M^2$ and $x_F$:
\bea
\label{M_xf_error}
M^2=\frac{\xi_2}{\xi_1} \left( m_1^2 +\vec{p}_{1\perp}^2  \right);
\nn
x_F= 2 \xi_2. 
\phantom{ \left( m_1^2 +\vec{p}_{1\perp}^2  \right) \, }
\eea
Hence, the variables (\ref{vary_xi}) do not coincide with the Bjorken
variables in the Drell-Yan scaling limit.

Contrary to the statement of \cite{error}, the relations (\ref{M_xf_error}) 
are not a dynamics effect, but an artifact, caused by the use of the 
alternative definitions (\ref{vary_xi}). 
At the same time,the factorization in the form (\ref{fact}), applying the 
usual $p_\perp$-dependent parton distribution functions, is not applicable 
in this system of reference, because the motion of the target parton is not confined 
to a light cone even in the high $S$ limit. 

\end{appendix}
%%%%%%%%%%%%%%%%%%%%%%%%%%%%%%%%%%%%%%%%%%%%%%%%%%%%%%%%%%%%%%%%%%%%%%%%%%

%%%%%%%%%%%%%%%%%%%%%%%%%%%% Bibliografy %%%%%%%%%%%%%%%%%%%%%%%%%%%%%%%%%

%
\end{document}